\documentclass[twocolumn,showpacs,showkeys,amsmath,amssymb,nofootinbib,superscriptaddress,floatfix]{revtex4}

\usepackage{graphicx}
\usepackage{subfigure}
\usepackage{bm}

\makeatletter
\newcommand\erfc{\mathop{\operator@font erfc}\nolimits}
\def\slashchar#1{\setbox0=\hbox{$#1$}
   \dimen0=\wd0 \setbox1=\hbox{/} \dimen1=\wd1
   \ifdim\dimen0>\dimen1 \rlap{\hbox to \dimen0{\hfil/\hfil}} #1
   \else  \rlap{\hbox to \dimen1{\hfil$#1$\hfil}} / \fi}

\newcommand{\lrd}{\raisebox{0.09em}{$\stackrel{\scriptstyle\leftharpoonup\hspace{-.7em}\rightharpoonup}{D} $}}
\newcommand{\ld}{\raisebox{0.09em}{$\stackrel{\scriptstyle\leftharpoonup}{D}$}}
\newcommand{\rd}{\raisebox{0.09em}{$\stackrel{\scriptstyle\rightharpoonup}{D}$}}

\begin{document}
 
\title{A note on the QCD evolution of generalized form factors}

\author{Wojciech Broniowski} 
\affiliation{The H. Niewodnicza\'nski Institute of Nuclear Physics,
  Polish Academy of Sciences, PL-31342 Krak\'ow, Poland}
\affiliation{Institute of Physics, Jan Kochanowski University,
  PL-25406~Kielce, Poland} 
\author{Enrique Ruiz Arriola}
\affiliation{Departamento de F\'{\i}sica
  At\'omica, Molecular y Nuclear, Universidad de Granada, E-18071 Granada, Spain}

\date{21 January 2009}

\begin{abstract}
Generalized form factors of hadrons are objects appearing in moments of the
generalized parton distributions. Their leading-order
DGLAP-ERBL QCD evolution is exceedingly simple and the solution is
given in terms of matrix triangular structures of linear equations
where the coefficients are the evolution ratios. We point out that this solution has
a practical importance in analyses where the generalized form factors
are basic objects, {\em e.g.}, the lattice-gauge studies or models. It also displays general features of their 
evolution.
\end{abstract}



\maketitle 

Generalized parton distributions (GPDs) (for notations and discussion
see the extensive reviews
\cite{Ji:1998pc,Radyushkin:2000uy,Goeke:2001tz,Bakulev:2000eb,Diehl:2003ny,Ji:2004gf,Belitsky:2005qn,Feldmann:2007zz,Boffi:2007yc}
and references therein) carry very rich information on the internal
structure of hadrons. In particular, moments of the GPDs in the
$X$ variable, according to the polynomiality feature, can be
written as polynomials in the $\xi$ variable, assuming the form
\begin{eqnarray}
\int_{-1}^1 \!\!\!\!\! dX\,X^{2j} \, {F}^{ns}(X,\xi,t) = 2\sum_{i=0}^j A_{2j+1,2i}(t) \xi^{2i}, \nonumber \\
\int_{-1}^1 \!\!\!\!\! dX\,X^{2j+1} \, {F}^{s}(X,\xi,t) = 2\sum_{i=0}^{j+1} A_{2j+2,2i}(t) \xi^{2i}, \label{poly}
\end{eqnarray}
with $j=0,1,\dots$. Here $X$ is the average fraction of the target's
momentum carried by the struck quark, and $\xi$ is the fraction of the
momentum along the light cone passed to the target. We use the
so-called symmetric notation, where $X \in [-1,1]$ and $\xi \in
[0,1]$.  Indices $ns$ and $s$ denote the non-singlet and singlet
distributions. The singlet GPD consists of the quark and gluon parts,
$F^s(X,\xi,t)=(F^{s,q}(X,\xi,t),X F^{s,G}(X,\xi,t)$).  The {\em
  generalized form factors} $A_{nk}$ (GFFs) for the quark operators
are the matrix elements between states $i$ and $f$ of the form (we
take the even-parity operator and a spin 0 target)
\begin{eqnarray}
 \label{eq:gff}
 && \langle {f(p')} | \overline{\psi}(0)\, \gamma^{\{\mu}\, i \lrd\/^{\mu_1}
  i \lrd\/^{\mu_2} \dots i \lrd\/^{\mu_{n-1}\}} \,\psi(0) |{i(p)}\rangle = \nonumber \\
 && \;\;\;\;2 P^{\{\mu}P^{\mu_1} \dots P^{\mu_{n-1}\}} A_{n0}(t)
    +  \\ && \;\;\;\;2\sum^n_{\substack{k=2\\\textrm{even}}}q^{\{\mu}
    q^{\mu_1} \dots q^{\mu_{k-1}}
    P^{\mu_{k}} \dots P^{\mu_{n-1}\}} \,2^{-k}A_{nk}(t), \nonumber
\end{eqnarray}
with $n=1,2,3\dots$, $k=0,2,\dots,n$, $P=(p+p')/2$, $q=p'-p$, $t=q^2$,
and $\psi$ denoting the quark field.  The symbol $\rd$ is the QCD
covariant derivative, $\lrd = \frac{1}{2} (\rd -\ld)$, and $\{\dots\}$
denote the symmetrization of indices and the subtraction of traces for
each pair of indices $\mu_1$ \dots $\mu_{n-1}$. The factor of $2^{-k}$
is conventional.  A similar expression can be written for the singlet
gluon GFFs \cite{Belitsky:2005qn}.

For the simplest case of a spin-0 target with positive charge such as
e.g. $\pi^+$, $A_{10} (t) $ is the charge form factor, while $A_{20}
(t)$ and $-A_{22} (t)$ are the quark components of the gravitational form factors
\cite{PhysRev.144.1250,Donoghue:1991qv}. 

The GPDs undergo the QCD evolution with a change of the
renormalization scale. Unlike the parton distribution functions (PDFs)
or the distribution amplitudes (DAs), it is non-trivial to pass to the
space of moments where the evolution is diagonal and then invert the
transformation.  This issue makes the case different and more
complicated already at LO from the case of the PDFs, where the Mellin
moments are used, or the DAs, where the Gegenbauer moments diagonalize
the evolution.  Theoretical tools have been developed to achieve the
task, such as the Shuvaev transformations \cite{Shuvaev:1999ce}, the
dual representation of the GPDs \cite{Polyakov:2002wz,Polyakov:2008aa}
or techniques based on the conformal moments complemented with the
Mellin-Barnes
transformations~\cite{Kumericki:2006xx,Kumericki:2007sa}. One may also
solve the DGLAP-ERBL equations numerically
\cite{GolecBiernat:1998ja,Martin:2008tm}.  Such approaches are needed,
if the whole GPD is demanded. Frequently, however, one is only
interested or has access to a limited number of GFFs.  The purpose of
this note is to point out that in fact the LO DGLAP-ERBL evolution of
the GFFs is inherently much simpler and straightforward to implement
in practice, without any need of complicated mathematical
transformations. The result discussed in this paper is implicitly
present in numerous works concerning the evolution of GPDs
(\cite{Belitsky:2005qn} and references therein), but nevertheless we
find it practical to present its explicit form, useful for those
dealing with the GFFs only and not the full GPDs.

Our starting point is the work of Kivel and Mankiewicz
\cite{Kivel:1999wa,Kivel:1999sk}, which elaborates the formalism of
Balitsky and Braun \cite{Balitsky:1987bk} on the QCD string operators
in the coordinate space.  We use Eqs.~(18) and (20) from
Ref.~\cite{Kivel:1999wa} for the LO DGLAP-ERBL evolution of
the GPDs from the scale $\mu_0$, where one assumes they are known, to
the scale $\mu$:
\begin{eqnarray}\label{eq:NS}
&&\!\!\!{\cal F}^{ns}(\beta,\xi,t;\mu^2) =
  \frac{1}{\sqrt{\pi}}\,\left(\frac{2}{\beta \xi}\right)^\frac{3}{2}\,
  \sum_{n=0}^\infty (-1)^n \left ( \frac{3}{2}+2n \right ) \times
  \nonumber \\ && 
L_{2n + 1} J_{\frac{3}{2}+2n} (\beta \xi) \int_0^1
  \!\! d\omega \,F^{ns}(\omega,\xi,t;\mu_0^2) \, C^\frac{3}{2}_{2
    n}\left (\frac{\omega}{\xi} \right). \nonumber \\ &&\!\!\!{\cal
    F}^s(\beta,\xi,t;\mu^2) =
  \frac{1}{\sqrt{\pi}}\,\left(\frac{2}{\beta \xi}\right)^\frac{3}{2}\,
  \sum_{n=0}^\infty (-1)^n \left ( \frac{5}{2}+2n \right )
  \times 
\nonumber \\ && 
L_{2 n + 2} J_{\frac{5}{2}+2n}
  (\beta \xi) \int_0^1 \!\! d\omega \,F^s(\omega,\xi,t;\mu_0^2) \,
  C^\frac{3}{2}_{2 n + 1}\left ( \frac{\omega}{\xi} \right), \label{eq:S} 
\end{eqnarray}
where
\begin{eqnarray}
\label{eq:curly} 
{\cal F}^{ns}(\beta,\xi,t;\mu^2)&=& \frac{2}{\pi}\!\int_{0}^1 \!\!
d\omega \,F^{ns}(\omega,\xi,t;\mu^2) \, \cos{\left(\omega \beta
  \right)} \, , \nonumber \\ {\cal F}^s(\beta,\xi,t;\mu^2) &=&
\frac{2}{\pi}\!\int_{0}^1 \!\! d\omega \,F^s(\omega,\xi,t;\mu^2) \,
\sin{\left(\omega \beta \right)}.
\end{eqnarray}
For the non-singlet case the quantity $L_{2n+1}$ denotes the evolution
ratio. For the singlet case $L_{2n+2}$ forms a 2-dimensional matrix in
the quark-gluon space. Explicitly,
\begin{eqnarray}
L_{2n+1}&=&\left ( \frac{\alpha(\mu^2)}{\alpha(\mu_0^2)}\right )^{\gamma_n^{qq}/(2\beta_0)}\!\!\!\!, \;
L_{2n+2}=\left ( \frac{\alpha(\mu^2)}{\alpha(\mu_0^2)}\right )^{\Gamma_n/(2\beta_0)}, \nonumber \\
\Gamma_n&=&\left ( \begin{array}{cc} \gamma_n^{qq} & \gamma_n^{qG} \\ \gamma_n^{Gq} & \gamma_n^{GG}  
\end{array} \right ),
\end{eqnarray}  
where the $\gamma$'s denote the appropriate anomalous dimensions.
%
Our further procedure is based on the
observation that ${\cal F} (\beta,\xi,t;\mu^2) $ is the generating
function of the the GFF's. We then expand
Eq.~(\ref{eq:NS}-\ref{eq:curly}) in $\beta$ around 0 using the
series expansions 
\begin{eqnarray}
J_m(z)&=&\sum_{l=0}^\infty \frac{(-1)^l}{l!(m+l)!}\left ( \frac{z}{2}\right )^{2l+m} , \\ 
C_m^{\lambda}(z) &=&\sum _{l=0}^{\left\lfloor \frac{m}{2}\right\rfloor } \frac{(-1)^l (2 z)^{m-2 l} (\lambda )_{m-l}}{l!
   (m-2 l)!},
\end{eqnarray}
where $(.)_n$ denotes the Pochhammer symbol, and finally use the
polynomiality property (\ref{poly}). As a result, polynomials in
$\beta$ and $\xi$ are obtained on both sides of
Eq.~(\ref{eq:NS},\ref{eq:S}). For subsequent values of the powers of
$\beta$ we compare the coefficients of powers of $\xi$.  As a result,
the equations for the form factors from Eq.~(\ref{poly},\ref{eq:gff}) follow immediately. We use
the short-hand notation $A_{nk}=A_{nk}(X,\xi,t;\mu)$ and
$A_{nk}^0=A_{nk}(X,\xi,t;\mu_0)$. For the non-singlet case
\begin{eqnarray}
&& A_{2k+1,2l}= k  \Gamma (2 k)\sum_{m=0}^k (4m+3)L_{2m+1} \sum_{j=k-l}^k \\ &&\frac{2^{2 (j-k)} (-1)^{m-j} 
\Gamma \left( j+m+\frac{3}{2}\right) A^0_{2 j+1,2 (j-k+l)}}{\Gamma (2 j+1) \Gamma (m-j+1) \Gamma (k-m+1)
   \Gamma \left( k+m+\frac{5}{2}\right)}, \nonumber
\end{eqnarray}
for $k=0,1,2,\dots$ and $l=0,1,\dots,k$, or, explicitly,
\begin{eqnarray}
A_{10}&=&L_1 A_{10}^0, \nonumber \\
A_{32}&=&\frac{1}{5}(L_1-L_3)A_{10}^0+L_3 A_{32}^0, \nonumber \\
A_{54}&=&\frac{1}{105}(9L_1-14L_3+5L_5)A_{10}^0 \nonumber \\&&+\frac{2}{3}(L_3-L_5)A_{32}^0+L_5 A_{54}^0, \nonumber \\
&\dots& \nonumber \\  
A_{30}&=&L_3 A_{30}^0, \nonumber \\
A_{52}&=&\frac{2}{3}(L_3-L_5)A_{30}^0+L_5 A_{52}^0, \nonumber \\
&\dots& \nonumber \\ 
A_{50}&=&L_5 A_{50}^0,\label{ev:ns}
\end{eqnarray}
where we have grouped the equations in the growing difference of the
indices $n$ and $i$ in $A_{ni}$. The ellipses denote equations with
$n\ge 5$.  Since $L_1=1$, the vector form factor, of course, does not
evolve. All other form factors in Eq.~(\ref{ev:ns}) change.  While the
{\it standard } form factors $A_{n0}$ retain their shape, {\em i.e.}
$A_{n0}(t)/A_{n0}(t=0)$ is not altered by the evolution, other genuine
{\it generalized} form factors involve linear combinations and both
their value at $t=0$ and their shape do change. Analogously, for the
singlet case
\begin{eqnarray}
&&A_{2k+2,2l}=\Gamma (2 k+2) \sum_{m=0}^k (4m+5)L_{2m+2} \sum_{j=k-l}^k \\ &&
\frac{2^{2 j-2 k-1} (-1)^{m-j} \Gamma \left( j+m+\frac{5}{2}\right) A^0_{2 (j+1),2 (j-k+l)}}{\Gamma (2 j+2) \Gamma (m-j+1) \Gamma
   (k-m+1) \Gamma \left( k+m+\frac{7}{2}\right)}, \nonumber
\end{eqnarray}
for $k=0,1,2,\dots$ and $l=0,1,\dots,k+1$, or, explicitly,
\begin{eqnarray}
A_{22}&=&L_2 A_{22}^0, \nonumber \\
A_{44}&=&\frac{3}{7}(L_2-L_4)A_{22}^0+L_4 A_{44}^0, \nonumber \\
A_{66}&=&\frac{5}{231}(11L_2-18L_4+7L_6)A_{22}^0 \nonumber \\&&+\frac{10}{11}(L_4-L_6)A_{44}^0+L_6 A_{66}^0, \nonumber \\
&\dots& \nonumber 
\end{eqnarray} 
\begin{eqnarray}
A_{20}&=&L_2 A_{20}^0, \nonumber \\
A_{42}&=&\frac{3}{7}(L_2-L_4)A_{20}^0+L_4 A_{42}^0, \nonumber \\
A_{64}&=&\frac{5}{231}(11L_2-18L_4+7L_6)A_{20}^0 \nonumber \\&&+\frac{10}{11}(L_4-L_6)A_{42}^0+L_6 A_{64}^0, \nonumber \\
&\dots& \nonumber \\ 
A_{40}&=&L_4 A_{40}^0,\nonumber \\
A_{62}&=&\frac{10}{11}(L_4-L_6)A_{40}^0+L_6 A_{62}^0, \nonumber \\
&\dots& \nonumber \\
A_{60}&=&L_6 A_{60}^0,\label{ev:s}
\end{eqnarray}
Note the identical structure of the first two groups in the above  equation. 
Again, the shape of the form factors $A_{n0}$ does not change.  The
sets of equations (\ref{ev:ns},\ref{ev:s}), although implicitly present 
in schemes involving the conformal moments, have not, to our knowledge, been 
written explicitly and their practical importance has not been recognized.
Since all quantities
on the right-hand side are known, from any practical point of view the
problem of the leading-order DGLAP-ERBL evolution of the GFFs is solved.  

Expressions for higher values of $n$ may be
obtained from the general expressions, however, it is the lowest form factors
which are most relevant, as they can be obtained in the Euclidean lattice
studies for the pion \cite{Brommel:2005ee,Brommel:PhD,Brommel:2007xd}
and the nucleon
\cite{Hagler:2003jd,Gockeler:2003jfa,Gockeler:2006zu,Hagler:2007xi}.
Equations (\ref{ev:ns},\ref{ev:s}) are useful for the evolution of higher
GFFs which eventually will be measured on the
lattice as the accuracy is increased, as well as for various model
calculations, where the results need to be evolved in order to compare
to the data \cite{Broniowski:2008hx}.

Sequences of equations in
(\ref{ev:ns},\ref{ev:s}) separated by ellipses form (infinite)
triangular matrix structures.  Mixing occurs between the $ns$ or $s$
form factors where the difference $n-i$ in $A_{ni}$ is fixed, for instance $A_{10}$, $A_{32}$,
$A_{54}$, {\em etc.} This corresponds, according to
Eq.~(\ref{eq:gff}), to the mixing of the $t$-channel states of the same
angular momentum. Indeed, $n-i$ is the number of Lorentz
indices of the $t$-channel momentum $q$.  These triangular matrix
equations may be diagonalized, yielding the combinations 
\begin{eqnarray}
A_{10}, \;\; A_{32}-\frac{1}{5}A_{10}, \;\; A_{54}-\frac{2}{3}A_{32}+\frac{1}{21}A_{10}, \label{eq:diago}
\end{eqnarray}
{\em etc.}, which evolve autonomously with $L_n$. The coefficients in the above combinations 
are proportional to the coefficients of the Gegenbauer polynomial $C^{3/2}_{n}(z)$. 
In fact, we are simply recovering the well known fact \cite{Belitsky:2005qn} that the conformal moments of the GPDs,
\begin{eqnarray}
\!\!\!\!\! \frac{\Gamma(3/2)\Gamma(n+1)}{2^{n+1}\Gamma(n+3/2)} \int_{-1}^1 \!\! dX \xi^n C_n^{3/2}(X/\xi) F^{ns,s}(X,\xi,t),
\end{eqnarray}
evolve autonomously at LO.
The point is, however, that the diagonalization of Eq.~(\ref{eq:diago}) is not necessary
for the evolution of the GFFs, as for any
practical purpose one can simply apply Eq.~(\ref{ev:ns},\ref{ev:s}).

Asymptotically, as $\mu^2 \to \infty$ we have (for the positive-parity operator) $L_1=1$ and $L_n \to 0$ for
$n>1$. Hence, in the non-singlet channel
\begin{eqnarray} 
A_{10}&\to &A_{10}^0, \; A_{32}\to \frac{1}{5}A_{10}^0, \; A_{54}\to
\frac{9}{105}A_{10}^0 \; \dots \nonumber \\ A_{30}&\to &
0, \; A_{52}\to 0, \; \dots \; A_{50}\to 0,\; \dots \label{asym1}
\end{eqnarray}
In the singlet channel we give the sum of the quark and gluon
components, whose second moment for the assumed vector case is related
to the momentum sum rule.  Denoting $S_{nk}=A_{nk}^q+A_{nk}^G$ we find
asymptotically
\begin{eqnarray}
S_{22}&\to &S_{22}^0, S_{44}\to \frac{3}{7}S_{22}^0, \; S_{66} \to \frac{5}{21}S_{22}^0 , \; \dots \nonumber \\ 
S_{20}&\to&S_{20}^0, \; S_{42}\to \frac{3}{7}S_{20}^0, \; S_{64}\to \frac{5}{21}S_{20}^0, \; \dots \nonumber \\ 
S_{40}&\to&0,\; S_{62}\to 0, \; \dots \; S_{60}\to 0, \dots \label{asym2}
\end{eqnarray}
Equations (\ref{asym1},\ref{asym2}) comply to the following LO asymptotic forms of the quark and gluon GPDs,
\begin{eqnarray}
\label{eq:asymptotics}
F^{ns}&\to& \theta(\xi^2-X^2)\frac{3}{2\xi} \left(1-\frac{X^2}{\xi^2}\right) A_{10}(t), \nonumber
\\
F^{s,q}&\to& \theta(\xi^2-X^2) (S_{20}(t)+ \xi^2 S_{22}(t)) \nonumber \\ 
&& \times \frac{15}{4 \xi^2} \frac{N_f}{4 C_F + N_f} \frac{X}{\xi}\left(1-\frac{X^2}{\xi^2}\right),\nonumber
\\
X F^{s,G}&\to& \theta(\xi^2-X^2)(S_{20}(t) + \xi^2 S_{22}(t)) \nonumber \\
&& \times \frac{15}{16 \xi} \frac{4C_F}{4C_F + N_f} \left(1-\frac{X^2}{\xi^2}\right)^2.
\end{eqnarray} 
The lowest form factors determine these expressions. Note,
however, that the gravitational form factors $S_{20}(t)$ and
$S_{22}(t))$ need not be equal.  Only for the special case
$S_{20}(t)=-S_{22}(t)=\theta(t)$ one recovers the typically written
form with the common factor $(1-\xi^2) \theta(t)$.  Since $S_{22}$
corresponds to the coupling of a scalar, and $S_{20}$ to the traceless
rank-2 tensor, there is no reason why $S_{20}(t)=S_{22}(t)$ should
hold. This issue is related to the lack of factorization of the
$t$-dependence in GPDs.

In conclusion, we state that the generalization of the result to other
channels is straightforward. For other probing operators one needs to
simply use appropriate anomalous dimensions. For various targets
(pion, nucleon) where different tensor couplings appear, one evolves
the form factors separately for the independent structures.

Gravitational and higher-order GFFs may be obtained from chiral quark
models of the GPDs for the pion
\cite{Polyakov:1999gs,Theussl:2002xp,Tiburzi:2002kr,Tiburzi:2002tq,Bissey:2003yr,Noguera:2005cc,Broniowski:2007si,Broniowski:2008hx}
and for the nucleon \cite{Goeke:2007fp,Goeke:2007fq}. A related
quantity, the pion-photon transition distribution amplitude
\cite{Pire:2004ie,Pire:2005ax,Lansberg:2006fv,Lansberg:2007bu}, has
also been obtained in quark models
Refs.~\cite{Tiburzi:2005nj,Broniowski:2007fs,Courtoy:2007vy,Courtoy:2008af,Kotko:2008gy}
and its moments undergo the QCD evolution in a similar way. Dynamical
calculations test the simplifying but {\it a priori} unjustified
assumptions made in many phenomenological studies. In particular, the
widely assumed factorization of the $t$-dependence is disproved. On
the other hand, the reference scale, $\mu_0$, turns out to be very low
in chiral quark models, around 320~MeV for the local models, and hence
QCD evolution to experimentally accessible scales implies a long
distance evolution. This is a case where the dilatation covariance may
prove crucial, as it ensures the integrability of the renorm-group
equations, hence the evolution path independence between two different
scales~\cite{RuizArriola:1998er}.

Actually, much of the explicit simplicity of the GFF evolution of Eq.~(\ref{ev:ns},\ref{ev:s}) is
linked to the LO approximation and the conformal invariance of the
evolution which represents faithfully the dilatation group,
a feature not automatically guaranteed in the current factorization
approaches at NLO, possibly inducing systematic errors. Renorm-group
improvement of the GPDs might be implemented as  previously done for
the DGLAP evolution of the PDFs~\cite{RuizArriola:1998er} as well as the
ERBL evolution of the PDAs~\cite{Bakulev:2005vw}. While these NLO
complications prevent writing down a handy analytic solution for the GFF
evolution, the problem can be reduced to a set of coupled differential
equations. It remains a tractable and much simpler
alternative when a reduced set of GFFs is available. 
See also Refs.~\cite{Kumericki:2006xx,Kumericki:2007sa}.

\medskip

We are grateful Lech Szymanowski for useful discussions. 

Research supported by the Polish Ministry of Science and Higher
Education, grants N202~034~32/0918 and N202~249235, Spanish DGI and
FEDER funds, grant FIS2008-01143, Junta de Andaluc{\'\i}a grant
FQM225-05, and EU Integrated Infrastructure Initiative Hadron Physics
Project contract RII3-CT-2004-506078.


\end{document}